\documentclass[aps,prc,showpacs,superscriptaddress,twocolumn,10pt]{revtex4-1}
\usepackage{graphicx}
\usepackage{dcolumn}
\usepackage{bm}
\usepackage{color}
\usepackage{comment}
\usepackage{ulem}

\begin{document}

\title{
A 4-dimensional Langevin approach to low-energy nuclear fission
of $^{236}$U}

\author{Chikako Ishizuka}
\email{chikako@lane.iir.titech.ac.jp}
\affiliation{
Laboratory for Advanced Nuclear Energy, Institute of Innovative Research, Tokyo Institute of Technology, Tokyo, 152-8550 Japan
}
\author{Mark D. Usang}
\affiliation{
Laboratory for Advanced Nuclear Energy, Institute of Innovative Research, Tokyo Institute of Technology, Tokyo, 152-8550 Japan
}
\affiliation{
Malaysian Nuclear Agency, Bangi, 43000 Malaysia
}
\author{Fedir A. Ivanyuk}
\affiliation{
Laboratory for Advanced Nuclear Energy, Institute of Innovative Research, Tokyo Institute of Technology, Tokyo, 152-8550 Japan
}
\affiliation{
Institute for Nuclear Research, Kiev, 03028 Ukraine
}
\author{Joachim A. Maruhn}
\affiliation{Institute for Theoretical Physics, Goethe University, Frankfurt am Main, 60323 Germany}
\author{Katsuhisa Nishio}
\affiliation{Advanced Science Research Center, Japan Atomic Energy Agency, Tokai, Ibaraki 319-1195 Japan}
\author{Satoshi Chiba}
\affiliation{
Laboratory for Advanced Nuclear Energy, Institute of Innovative Research, Tokyo Institute of Technology, Tokyo, 152-8550 Japan
}
\affiliation{
National Astronomical Observatory of Japan, Tokyo, 181-8588 Japan
}

\date{\today}

\begin{abstract}
We developed a four-dimensional Langevin model
which can treat the deformation 
of each fragment independently
and applied it to low energy fission of $^{236}$U, the compound system of 
the reaction $n + ^{235}$U.
The potential energy is calculated with the deformed two-center Woods-Saxon
(TCWS) and the Nilsson type potential with the microscopic energy
corrections following the Strutinsky method and BCS pairing.
The transport coefficients are calculated by macroscopic prescriptions.
It turned out that the deformation for the light and heavy fragments behaves differently, showing a sawtooth structure similar to
that of the neutron multiplicities of the individual fragments $\nu(A)$. Furthermore, the measured
total kinetic energy $TKE(A)$ and its standard deviation 
are reproduced fairly well by the 4D Langevin model based on the TCWS potential in addition to the  fission fragment mass distributions. 
The developed model allows a multi-parametric correlation analysis 
among, e.g., the three key fission observables, mass, TKE, and neutron multiplicity, which should be essential to elucidate several
long-standing open problems in fission such as the sharing of the excitation energy between the fragments.
\end{abstract}


\maketitle
\section{Introduction}
Nuclear fission is a unique 
large-amplitude collective motion of 
nuclear matter which should be described in principle as a quantum many-body system.
Predicting fission observables with high accuracy by studying the underlying properties of nuclear matter has been one of the challenging topics in nuclear physics, especially for low-energy fission where microscopic properties associated with the shell structure play an essential role.
For this goal, various theoretical models have been proposed, see for example the recent review article \cite{Bertsch2015}.
Among the experimental data, the fission fragment mass distribution (FFMD), 
the total kinetic energy (TKE),
and the prompt neutron multiplicity $\nu(A)$, all as functions of mass number of fission fragments, are the most important fission observables. They are largely connected to the configuration at the scission point, characterized by the fragment mass-asymmetry, the Coulomb repulsion energy, and the deformation energies of both fragments.
Furthermore, they are strongly correlated with each other under energy conservation through the fission process. It is well known that $\nu(A)$ shows a so-called \textit{sawtooth} structure and has a mirror-asymmetry around symmetric fission (e.g. \cite{Nishio+1998}), indicating an independence of the deformation of both fragments. A reliable fission theory for prediction of these key observables should thus include at least four shape parameters: the mass-asymmetry, charge-center distance (elongation) and the deformation of 
each fragment.

Quite recently the concept of the Brownian shape motion was introduced in nuclear fission, which demonstrated 
high-predictive power of calculating FFMD \cite{Moeller+2001,Randrup+2011} by performing random walks on a five-dimensional potential energy surface.
In this model, however, important features of nuclear dynamics such as energy dissipation cannot be 
treated, due to the assumption of overdamped motion. Thus several important quantities in fission such as the prescission kinetic energy (PKE) \cite{Usang+2016} and the fission time scale \cite{Jacquet09} 
have not been considered at the moment
in this framework. 

The fluctuation-dissipation model (the Langevin formula) can calculate
the time evolution of energies associated with the collective motion as well as their dissipation into intrinsic excitation energy,
thus the prescission kinetic energy and intrinsic excitation energy at the scission point
are calculated on the same footing.
The model can also determine the fission time scale, which is not the case
for the Random-walk method~\cite{Moeller+2001}.
At present there are several groups which use the Langevin approach for the description of the fission process~\cite{Usang+2016,Asano+2004,Mazurek+2015,Pahlavani+2015,Eslamizadeh+2017,Sadhukhan+2016}.
Due to the difficulty of the calculation
of the multi-dimensional transport coefficients used in the Langevin equations,
and due to the lack of sufficient resources for numerical calculations  the number of shape parameters in most cases is restricted to 2 or 3 collective variables. The only exception is a very recent work ~\cite{LAUR23344} where the mass and TKE distributions of fission fragments were calculated within the 5-dimensional Langevin approach with macroscopic transport coefficients.
Their calculation starts at a point outside the saddle due to, again, a reason of the huge computation time required.

In the case of 3-dimensional models, the three parameters 
typically used in
the Langevin equations are elongation, mass-asymmetry, and fragment deformation. Sometimes the neck-radius is chosen instead of fragment deformation. In all these cases the deformations of both fragments were confined to be identical.  
On the other hand, the low-energy fission data show a behavior which cannot be treated properly by the 3D model as explained above. 
One of the typical examples is the $\nu(A)$ of the fission of actinide nuclei, 
showing the different deformations of both fragments at the scission point.
This was solved in this work by developing a 4D-Langevin model
which can treat the deformation of each fragment independently.  
In addition, our Langevin trajectories start from a point 
inside the saddle where the compound 
system stays a long time and reaches to a state of quasi-equilibrium, a condition which must be satisfied 
implicitly for the concept of the Langevin theory to be valid.

This paper is organized as follows.  In Chapter II,  the two-center shell model parametrization
to express nuclear shapes appearing in fission is explained, and collective variables we treat are defined.
In Chapter III, a new potential formulation, the two-center Woods-Saxon model, is introduced.  
In Chapter IV, the Langevin equations are explained with supplementary theorems and formulae necessary 
to carry out the calculation.
In Chapter V, results of numerical calculations are shown and new insights into the dynamical aspects of 
fission, obtained by the present 4D approach, will be described.  Chapter VI is devoted to the summary of 
this paper.

\section{The two-center shell model}
\begin{figure}[h]
\includegraphics[width=0.7\columnwidth]{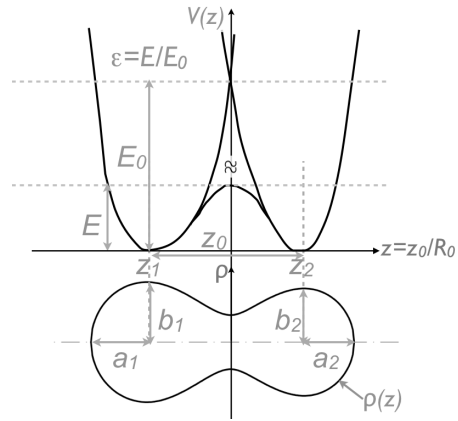}
\caption{(color online) The lower figure shows a snapshot of the configuration of a $^{236}$U nucleus calculated by the TCSM.
The upper figure is the corresponding potential shape. Two harmonic oscillator potentials are smoothly
connected around the elongated neck. The neck parameter $\epsilon$ is defined as the ratio
of the intercept of the harmonic oscillator potentials and that of the connecting function. 
} 
\label{Fig-TCSM}
\end{figure}

In the present paper we use the two-center 
{shell model (TCSM)} 
parametrization of nuclear shape suggested by Maruhn and Greiner \cite{Maruhn+1972}
to express a set of nuclear shapes appearing in fission.
In this model the mean-field potential includes the central part $V(\rho, z)$, as well as
$\boldsymbol{l}\boldsymbol{s}$ and $\boldsymbol{l^2}$ terms.
The central part potential $V(\rho, z)$ in the TCSM consists of two oscillator potentials smoothly joined together by a fourth-order polynomial in $z$, see Eq. (\ref{v_tcsm}) and Fig. \ref{Fig-TCSM}.  It is defined as
\begin{equation}\label{v_tcsm}
  V(\rho,z) = \left\{
 \begin{array}{lr}
 \frac{1}{2}m \omega_{z_1}^2 (z-z_1)^2 + \frac{1}{2}m \omega_{\rho_1}^2 \rho^2;& z \leq z_1 \\
 \frac{1}{2}m \omega_{z_1}^2 (z-z_1)^2 f_1(z,z_1) + \\
 \frac{1}{2}m \omega_{\rho_1}^2 \rho^2 f_2(z,z_1);&  z_1 \leq z \leq 0 \\
 \frac{1}{2}m \omega_{z_2}^2 (z-z_2)^2 f_1(z,z_2) + \\
 \frac{1}{2}m \omega_{\rho_2}^2 \rho^2 f_2(z,z_2);&  0 \leq z \leq z_2 \\
 \frac{1}{2}m \omega_{z_2}^2 (z-z_2)^2 + \frac{1}{2}m \omega_{\rho_2}^2 \rho^2;& z_2 \leq z
 \end{array}\right.
\end{equation}
with the quadratic functions in $z$
\begin{eqnarray}\label{f1f2}
f_1(z,z_i) &=& 1 + c_i (z-z_i) + d_i (z-z_i)^2\ ,\nonumber \\
f_2(z,z_i) &=& 1 + g_i (z-z_i)^2,~~(i=1,2) \ , 
\end{eqnarray}
where $c_i$, $d_i$, and $g_i$ are constants.
The shape of the nuclear surface in the TCSM is fixed by the requirement that at the surface $\rho=\rho(z)$ the potential  $V(\rho(z),z)$ is constant.

The central potential given in Eq.~(\ref{v_tcsm}) contains 12 parameters. By imposing the conditions that
$V(\rho, z)$ and its $z$-derivative are continuous at $z=\{z_1, 0, z_2\}$ 
and the volume conservation,
the number of parameters is reduced to 5. Two of them are the elongation parameter $z_0\equiv z_2-z_1$, the mass asymmetry $\alpha={(V_1 - V_2)}/{(V_1 + V_2)}$ ($V_1$ and $V_2$ are the volume 
of the left- and right-side
from $z=0$).
The ratios of oscillator frequencies $\omega_{\rho_i}/\omega_{z_i} (i=1,2)$ were expressed in terms of another two deformation parameters $\delta_i$~\cite{suek74,iwam76},
\begin{equation}\label{deltas}
\frac{\omega_{\rho_i}}{\omega_{z_i}}= \frac{3+\delta_i}{3-2 \delta_i}\,.
\end{equation} 
The ratios $\omega_{\rho_i}/\omega_{z_i}$ define the deformation of the left and right oscillator potentials and, thus, the deformation of the outer  ($z\leq z_1$ or $z_2\leq z$)  spheroidal part of the fragments, since $\omega_{\rho_i}/\omega_{z_i}=a_i/b_i$, where $a_i$ and $b_i$ are the semiaxes in the $z$ and $\rho$ direction, respectively, see Fig.~\ref{Fig-TCSM}. The deformation of the inner part of the nucleus ($z_1\leq z \leq z_2$) depends on all 5 deformation parameters. Therefore, in general, it does not mean that the fragment as a whole is prolate if $\delta_i$ is positive, or oblate when $\delta_i$ is negative. 

The fifth parameter $\epsilon$ is 
defined as the ratio of the
potential height $E$ at $z=0$ to the value $E_{0}$ of the left and right harmonic oscillator potentials at $z=0$, see Fig~\ref{Fig-TCSM}.
In our present and previous calculations~\cite{Usang+2016}, we have fixed $\epsilon=0.35$. This value leads to shapes that are very close to the so called optimal shapes \cite{ivchar} - the shapes that correspond to the lowest liquid drop energy at fixed elongation and mass asymmetry.
All parameters that appear in Eqs. (\ref{v_tcsm}) and (\ref{f1f2}) can be expressed in terms of the above 5 parameters.

In order to make the Langevin model realistic and widen the fission observables to be studied, we use in the present work a 4-dimensional TCSM shape parametrization such  that the fission fragments can have  independent deformations.
The set of deformation parameters is: $\{q_i\}\equiv \{z_0/R_0,\delta_1,\delta_2 ,\alpha\}$,
where $z_0/R_0$, $\delta_1$, $\delta_2$, and $\alpha$ are
the elongation of the compound nucleus, the deformation of both outer parts of the nucleus, and the mass asymmetry, while the 3D model had the restriction of $\delta_1=\delta_2=\delta$.
$R_0$ is the radius of a spherical compound nucleus.
The neck parameter $\epsilon$, being fixed at $0.35$ in the present calculation,
can be an additional shape parameter for the future development to the 5D Langevin model.

\begin{figure}[h]
\includegraphics[width=0.8\columnwidth]{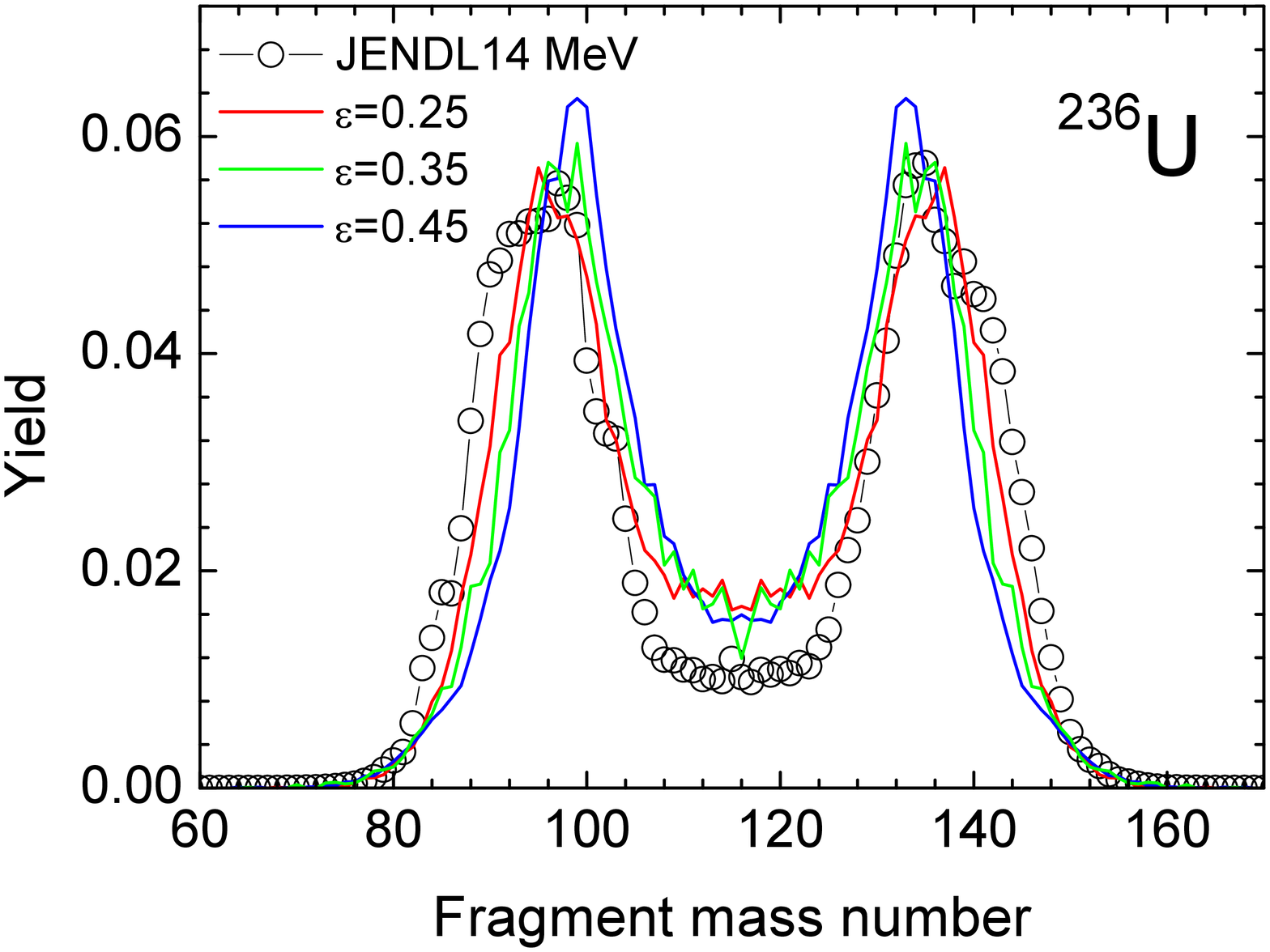}
\caption{(color online) 
Comparison of 4-dimensional calculations of the mass distribution of fission fragments with different neck parameters $\epsilon$ for $^{236}$U at the excitation energy of 20 MeV.  The experimental data are taken 
from JENDL/FPY-2011 data library\cite{jendl} for 14 MeV neutrons impinging on $^{235}$U.
}
\label{Figs:FFMD_epsilon}
\end{figure}

Please note that within the TCSM the neck radius depends not only on $\epsilon$ but on all other four parameters. 
Even in case of fixed $\epsilon$, the neck radius varies in a very broad region due to the variation of the other parameters. We have checked that the variation of $\epsilon$ within reasonable limits does not affect the calculated results noticeably as shown in Fig.~\ref{Figs:FFMD_epsilon}.

Throughout this paper, whenever we compare the calculated values with experimental data, our calculations are performed for $^{236}$U at the same excitation energy which is populated as a compound nucleus in the neutron-induced fission on $^{235}$U, for which experimental data are most abundant.  Furthermore, we ignored
the contribution of multi-chance fission by constraining our analysis to the low-energy region.  Still, we notice that effects of multi-chance fission should exist in some cases we studied.  Such effects can be described by combining the Langevin calculation with statistical Hauser-Feshbach theory.  Such an analysis will be an important future subject, but we did not attempt to do that in the present work since we wish to elucidate how the newly developed 4D Langevin model can describe the fundamental aspects of low-energy nuclear fission
without complication coming from the other effects.

\section{The potential energy}
\begin{figure}[ht]
\includegraphics[width=0.8\columnwidth]{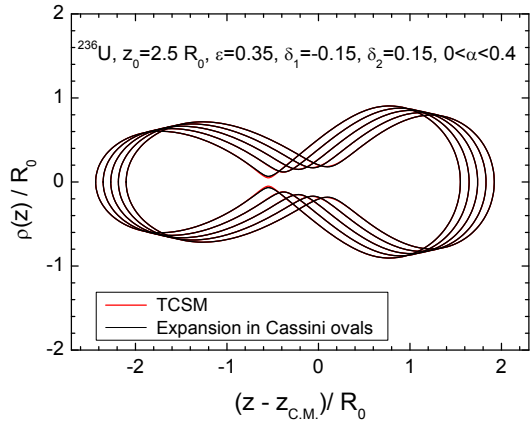}
\caption{(color online) An example of the expansion of the TCSM shapes close to the scission point in a series of Cassini ovaloids (TCWS).} 
\label{cass-exp}
\end{figure}

It turns out that the extension 
of the number of dynamical variables within the TCSM framework, on which our previous 3D model was based, is not sufficient to reproduce the experimental results. Consequently we have modified also the mean-field potential. Instead of the Nilsson type of potential of the TCSM we used a
more realistic finite-depth Woods-Saxon (TCWS) potential. For this the shape function $\rho(z)$ of the TCSM was expanded in a series of Cassini ovaloids. 
In total, 20 deformation parameters were taken into account to describe closely enough the nuclear shape given by the TCSM, as shown in Fig.~\ref{cass-exp}.

For the shape given by an expansion in Cassini ovaloids the two-center deformed Woods-Saxon 
approach~\cite{Pashkevich+2008} was used in order to calculate the single-particle energies and shell corrections ~\cite{Strutinsky+1967,Strutinsky+1968,Brack+1972}.
The parameters of the Woods-Saxon potential in Ref.~\cite{Pashkevich+2008} were used.
In the macroscopic-microscopic model \cite{Strutinsky+1967,Strutinsky+1968,Brack+1972}, the energy correction originating from the shell structure in a nucleus is added to the classical macroscopic potential energy. Thus the macroscopic potential energy $U(q,T)$ can be expressed as 
\begin{equation}\label{upot}
U(q,T) = E^{Macro}_{def}(q) + \delta E(q,T)\ .
\end{equation}
The macroscopic part of the potential energy, $E^{Macro}_{def}$, was calculated within the finite-range liquid drop model \cite{Krappe+1979}. The temperature (excitation) dependence of the shell corrections $\delta E$ was estimated by the Ignatyuk prescription \cite{Ignatyuk+1975} with the damping energy $E_d=20$ MeV,
\begin{equation}\label{damp}
\delta E(q,T)=\delta E(q,T=0)\cdot e^{-E_x/E_d}\,,
\end{equation}
where $E_x$ is the excitation energy, see Eq.~(\ref{internalEnergy}) below.

The shell energy $\delta E$  contains the  contributions from the shell effects in total single-particle energy and in the pairing energy,
\begin{equation}
\label{shell-pair-corr}
\delta E(q, T=0) =\sum_{n,p} \left( \delta E_{shell}^{(n,p)}(q) + \delta E_{pair}^{(n,p)} (q) \right) .
\end{equation}
The $\delta E_{shell}$ and $\delta E_{pair}$ were calculated by the BCS approximation and Strutinsky prescription~\cite{Strutinsky+1967,Strutinsky+1968,Brack+1972} from the single-particle energies obtained with TCSM or TCWS shell models.

We consider in present work fission process at low excitation energies. The corresponding temperatures  do not exceed $1$ MeV. For such temperatures the damping of shell effects is not so large and it was neglected. So, the calculations were done with full shell effects taken into account.

\section{The langevin approach}
The Langevin equation is written as follows using the shape coordinates $q_i$ and their
conjugate momenta $p_i$.
\begin{eqnarray}
\dot{q_i} &=& m_{ij}^{-1}p_j  \\
\dot{p_i} &=& - \frac{\partial U}{\partial q_i} - \frac{1}{2}\frac{\partial m_{jk}^{-1}}{\partial q_i}p_jp_k
- \gamma_{ij}m_{jk}^{-1}p_k + g_{ij}R_j(t) \  .
\end{eqnarray}
The quantities, $m_{ij}$, $\gamma_{ij}$, and $g_{ij}R_{j}$, correspond to the inertial mass tensor, the friction tensor,
and the random force, respectively.
For the transport coefficients,
we adopt
the Werner-Wheeler approximation~\cite{Davis+1976} for the mass tensor $m_{ij}$
and the wall and window model~\cite{Blocki+1978} ($k_s=0.27$) for the friction $\gamma _{ij}$.
The random force $g_{ij}R_{j}(t)$ is the product of white noise $R_{j} (t)$ and the temperature dependent strength factors $g_{ij}$. The factors $g_{ij}$ are related to the temperature and friction tensor via the modified Einstein relation \cite{Hofmann+1998},
\begin{equation}
g_{ik} g_{kj} = T^*\gamma_{ij} \ ,\qquad {\rm with}\,\,\,
T^*=\frac{\hbar\varpi}{2}\coth{\frac{\hbar\varpi}{2T}}\,,
\nonumber
\end{equation}
where $T^*$ is the effective temperature.  The parameter $\varpi$ is
the local frequency of collective motion \cite{Hofmann+1998}.
The minimum of $T^*$ is given by ${\hbar\varpi}/{2}$, which 
corresponds to the zero point energy of oscillators
forming the heat bath. 
Based on the pioneering works \cite{Bohr+1939,Hill+1952}, we estimated the zero point energy 
as 1 MeV, which lies in the middle of the corresponding quantities for various modes 0.45 to 2.23 MeV
estimated in Ref.~\cite{Hill+1952}.  
The temperature $T$ in this context is related to the initial excitation energy $E_x$ and the internal energy $E_{int}$ by,
\begin{equation}
\label{internalEnergy}
\nonumber
E_{int}=E_x-\frac{1}{2}\left(m^{-1} \right)_{ij}p_{i} p_{j} - U(q,T=0)=aT^2\,,
\end{equation}
where $a$ is the level density parameter~\cite{Toke+1981}.

\begin{figure}[ht]
\includegraphics[width=8cm]{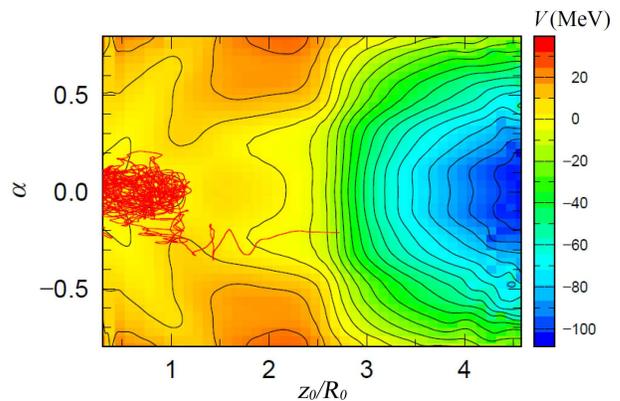}
\caption{(color online) A trajectory in the 4D-Langevin model (the red solid line) is shown on the color map of the potential energy surface
for $^{236}$U.
}
\label{Fig:trajectory}
\end{figure}
We started the Langevin calculation
as far inside the saddle as possible
in order to account for the stochastic fluctuation 
in an equilibrated medium of nuclear collective motion inside the saddle. 
 As shown in Fig.~\ref{Fig:trajectory} the trajectory stays inside the saddle for a long time, especially in the region near the potential minimum, 
before it gets over the saddle point of the potential energy surface. Thus, the distribution of trajectories at the saddle in mass asymmetry, kinetic and excitation energy 
emerges as a result of fluctuating motion inside the saddle. 
 Beyond the saddle the trajectory falls into the potential valley and reaches the scission point 
quickly.
If we started the calculation from the top of (or outside) the saddle, we would lose the 
stochastic nature of Langevin trajectories on the way from 
potential energy minima  
till the saddle point,
which are the essential features in the Langevin model
based on the fluctuation-dissipation dynamics.

Initially, the momenta $p_i$ were set to zero, and Langevin motions were initiated by the conservative and random forces. Such calculations are continued until the trajectories reach the scission points, which were defined as the points in deformation space where the neck radius becomes zero.
Using such a 4D-model, we have calculated the 
fission fragment mass distribution (FFMD),
total kinetic energy (TKE),
and its standard deviation $\sigma_{TKE}$
for $^{236}$U as a compound system of
neutron-induced fission of $^{235}$U.

\section{Numerical results}
\begin{figure}[htp]
\includegraphics[width=8cm]{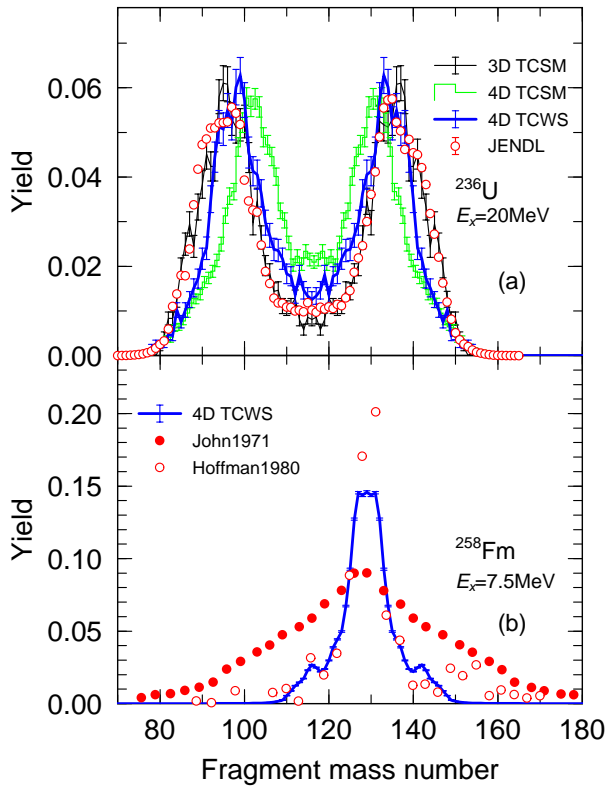}
\caption{
(color online) 
Panel (a): Mass distribution of fission fragments for the fission of 
$^{236}$U at an excitation energy of 20 MeV.  The 3D and 4D Langevin calculations (histograms) are compared with the 
experimental information given in the JENDL/FPY-2011 data library\cite{jendl} for n + $^{235}$U at 14 MeV.
Panel (b): Mass distribution of fission fragments for the fission of $^{258}$Fm at the excitation energy of 7.5 MeV is plotted with the experimental data for $^{258}$Fm spontaneous fission~\cite{Hoffman+1980} 
(red open circles) and the post-neutron FFMD of n$_{th}$+$^{257}$Fm fission~\cite{John+1971} (red filled circles). 
}
\label{Fig-01}
\end{figure}

The FFMDs for $^{236}$U and $^{258}$Fm are shown in Fig.~\ref{Fig-01}.
In both cases, we renormalized data in such a way that the total area becomes 2.
In Fig.~\ref{Fig-01} (a) we show the 
FFMD for $^{236}$U at $E_x=20$~MeV calculated with the 
3D and 4D Langevin models.
One can see that the use of the finite-depth TCWS potential (blue) reproduces the experimental peak positions and their widths better than the infinite-depth TCSM potential (green).
For comparison, results of the 3D Langevin model using the TCSM~\cite{Usang+2016} is shown
(black).
Both of the 3D TCSM model and the 4D TCWS model can reproduce well the whole structure of the experimental FFMDs. The peak widths in the 3D TCSM model is broader than in the 4D TCWS model. 
This is because of the difference of the 
potential energy surfaces between these models.
For symmetric components around $A=118$, the potential energy has the minimum at 
$\delta_1=\delta_2$ both in the 3D and 4D models.
On the other hand, for asymmetric components, 
the minimal potential energy can be lower in the 4D model than that in the 3D model
due to the additional degree of freedom ($\delta_1\neq\delta_2$). As a result,
the depths of potential valleys giving the two peaks become deeper in the 4D model
than that in the 3D model. It provides narrower peak widths in the 4D model.
However, it will become clear that the 4D model is suitable to describe the dynamical features
of fission better than the 3D model in a comprehensive manner due to its advantages of the 4D model to be shown below.

Fig.~\ref{Fig-01} (b)
shows the FFMD for $^{258}$Fm fission
with the excitation energy $E_x=7.5$ MeV and the full shell correction (no shell damping) as reference. 
The energy $E_x=7.5$ MeV, corresponds to the $^{258}$Fm spontaneous fission, because 
the fission barrier height of our model is about 7 MeV in this case.
In Fig.~\ref{Fig-01} (b), we compare our $E_x=7.5$ MeV result with $^{258}$Fm spontaneous fission data~\cite{Hoffman+1980} and
n$_{th}$+$^{257}$Fm fission data~\cite{John+1971}.
A strong single peak component can be seen in both experimental data.
However, please note that the thermal fission data~\cite{John+1971} was measured after prompt neutron emission.
According to their paper, a triple-humped FFMD was produced after their neutron correction although they also mentioned that the neutron correction has considerable uncertainty.

Thus, the finite-depth potential plays an essential role
to produce mass peaks at the right positions,
although we may need the improvement of the transport coefficients 
for more accurate FFMDs.
Further investigation of FFMDs of actinides will be discussed in 
a forthcoming paper. 

\begin{figure}[ht]
\includegraphics[width=8cm]{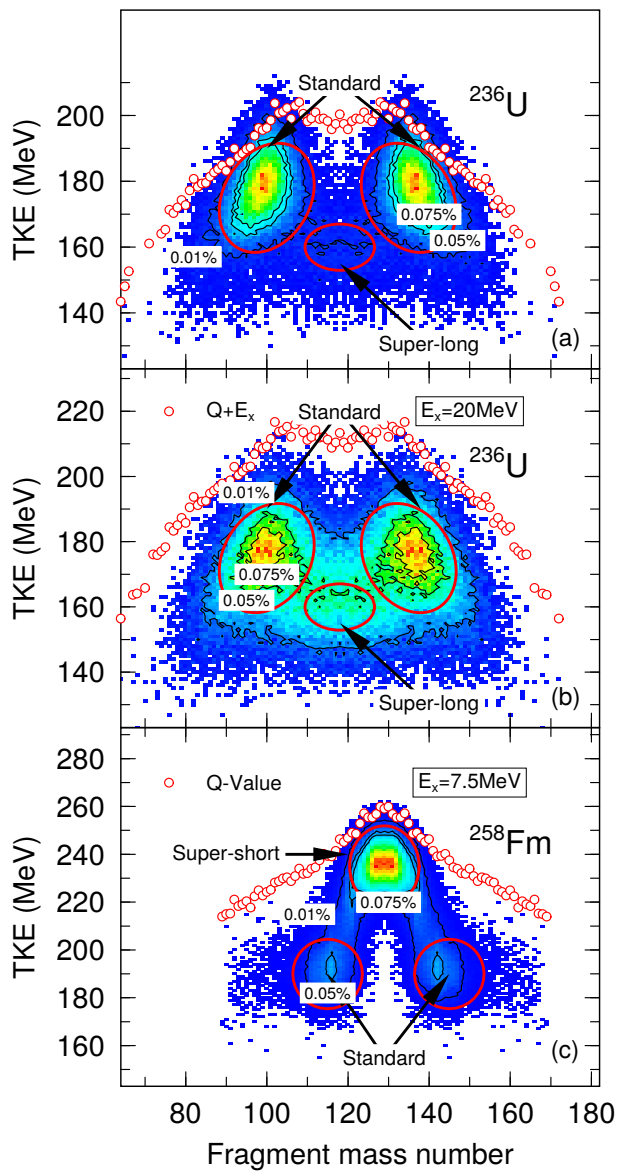}
\caption{\label{Fig:TKE-Q}
(color online) 
Calculated fission events on the mass-TKE plane for $^{236}$U-fission at $E_x=7$MeV (panel (a)) and $E_x=20$MeV(panel (b)).
Number of events are shown by different colors, increases from blue to red.
The locations of the standard and superlong modes in Brosa's terminology (see text) are shown by ellipses. 
The upper limit of the TKE, $Q + E_x$, from the mass database is shown by open circles~\cite{Audi+2003}.  
The same plot as panel (a) and (b) but in the case of $^{258}$Fm at $E_x=7.5$ MeV is shown in panel (c) for reference.} 
\end{figure}

\begin{figure}[h]
\includegraphics[width=8cm]{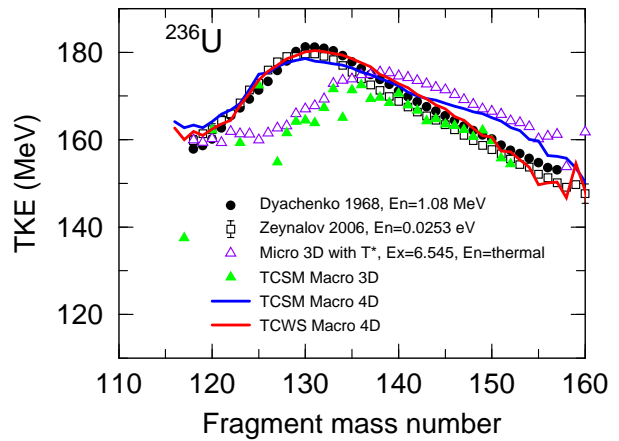}
\caption{\label{Fig-05}
(color online) TKE distributions, 
calculated by 3D and 4D Langevin models, 
for $^{236}$U-fission at $E_x=7$ MeV (corresponding to thermal neutron induced fission of $^{235}$U) are shown in comparison with the experimental data~\cite{zeynalov,exdepend_u235}.}
\end{figure}

As mentioned earlier, 
other models such as the random walk on a 5D potential surface~\cite{Moeller+2001},
also reproduce the mass distribution rather accurately.
Compared to such models, the advantage of the present approach  
lies in the fact that it can give a prediction of the TKE including the effects of prescission dynamics.
%
%
Our previous 3D model~\cite{Usang+2016},
which gives good agreement with the experimental FFMDs (Fig.~\ref{Fig-01} (a)), explains the qualitative behavior of the TKE, but it is not able 
to reproduce quantitatively the experimental values. In the current 4D model, both the TKE and its standard deviation agree
with the  experimental data, qualitatively and quantitatively as shown in
Figs.~\ref{Fig:TKE-Q},~\ref{Fig-05} and Table~\ref{Table-01}
(see also discussion below).
It is clear from Fig.~6 that the fission events are widely distributed around the average
TKE value. The average value of TKE is well below the upper limit of the TKE, $Q+E_x$. However,
we have noticed that some events give TKE which are larger than $Q+E_x$, even though
we checked that the constraint from the energy conservation was satisfied.

The $Q$ value was calculated with the assumption that the charge-to-mass ratio was conserved before and after scission, using the values of mass excess obtained from
Reference Input Parameter Library (RIPL-3) that provides
either experimental or recommended mass data~\cite{Audi+2003}.
For the case that we cannot find both experimental and recommended
mass data, we refer to the theoretical mass data in RIPL-3.
Discrepancy of nuclear masses given in RIPL-3 and those given by the TCWS could give rise to events
having TKE above the $Q+E_x$. 
This cannot be avoided since our TCWS is not optimized to reproduce
nuclear masses.
Apart from the definition of the $Q+E_x$ in our model,
in the Langevin equations, which on each integration step add (or subtract) some amount of kinetic energy, the additional energy gained from the random force accidentally can be large and can exceed the $Q+E_x$ at the final step of the calculation. In such cases the local intrinsic energy defined by Eq. (7) is negative and we put T=0.

%
%
%
In Brosa model~\cite{Brosa+1990} the nuclear scission process can be interpreted in terms of 
several fission modes, namely, the standard modes, the super-long modes and the super-short modes.
In Fig.~\ref{Fig:TKE-Q} (a) and (b), indeed, we can see the contributions from the standard and super-long modes in the mass-TKE distribution of $^{236}$U. 
The standard modes are dominant at $E_x=7$~MeV as shown in Fig.~\ref{Fig:TKE-Q} (a),
while
more super-long components appeared at higher excitation energy as shown in Fig.~\ref{Fig:TKE-Q} (b). 
In Fig.~\ref{Fig:TKE-Q} (c), we also show the TKEs for $^{258}$Fm at $E_x=7.5$~MeV as reference.
In $^{258}$Fm case, we can see that the super-short mode is dominant.
In our 3D Langevin study for Fm region~\cite{Usang+2018},
the microscopic transport coefficients is necessary to produce the TKE quantitatively.
However, the current 4D model can reproduce the averaged TKE values of the standard modes and the super-short mode
not only qualitatively but also quantitatively even with the macroscopic transport coefficients.

In Fig.~\ref{Fig-05}, 
we compare the calculated TKE values averaged at each fragment mass with the experimental data.  The calculations were done for compound $^{236}$U while the experimental data are for the neutron induced fission of $^{235}$U at incident energies of 0.0253 eV and 1.08 MeV.
A remarkable agreement is seen for the TKE distribution 
between the 4D calculations (the red solid line for the TCWS case, the blue solid line for the TCSM case) and two sets of experimental data.  
The two sets of 3D-models shown by triangles, one with microscopic transport coefficients, the other with macroscopic ones, cannot reproduce the data well.
Considering the agreement with the data for the mass and TKE distributions simultaneously, we can conclude that the 4D Langevin models are superior to the 3D ones.  Between the 4D models, 
on the other hand,
the agreement with the data is better in the TCWS than in the TCSM. 

The 4D model describes well not only the TKE but also its standard deviation $\sigma_{\rm TKE}$ 
in its dependence on the neutron energy $E_n$  
impinging on $^{235}$U, therefore, excitation energy of $^{236}$U, as tabulated in Table~\ref{Table-01}. 
It is known from experiments that the standard deviation of the TKE is almost constant, about 11 MeV, as a function of the neutron energy $E_n$.
The magnitude
of $\sigma_{\rm TKE}$ 
is also improved by the 4D models compared to the 3D ones.
\begin{table}[hb]
\caption{\label{Table-01} Standard deviation of the TKE of fission fragments, $\sigma_{TKE}$, for neutron induced fission on $^{235}$U (exp.) or excited $^{236}$U (calc.) corresponding to the same excitation energy.}
\begin{ruledtabular}
\begin{tabular}{lcccc}
\multicolumn{5}{c}{$\sigma_{TKE}$ [MeV]}\\
\textrm{$E_n$ [MeV]}&
\textrm{Present 4D}&
\textrm{3D}&
\textrm{Pre-n~\cite{Duke+2015}} &
\textrm{Post-n~\cite{Duke+2015}}\\
\colrule
0.5($E_x=7$)& 9.65 &6.19&10.65&10.85\\
3.5($E_x=10$)& 10.54 &6.63&10.60&10.83\\
5.5($E_x=12$)& 10.82 &7.43&10.83&10.99\\
8.5($E_x=15$)& 11.35 &8.38&10.90&11.09\\
13.5($E_x=20$)& 11.72 &9.54&11.18&11.44\\
 \end{tabular}
 \end{ruledtabular}
 \end{table}

Hereafter we will concentrate on the TCWS case, because the 4D Langevin approach with the TCWS potential reproduces the data better than the TCSM.
 The remarkable agreement of the TKE
shown in Fig.~\ref{Fig-05} and its standard deviation  given in Table \ref{Table-01} indicates that the 
prescission dynamics and the nuclear shape at scission is described well by the 4D Langevin calculation based on the TCWS potential since the 
TKE is the sum of the prescission kinetic energy and 
the Coulomb repulsion energy between the nascent fragments at the scission.

\begin{figure}[htb]
\includegraphics[width=8cm]{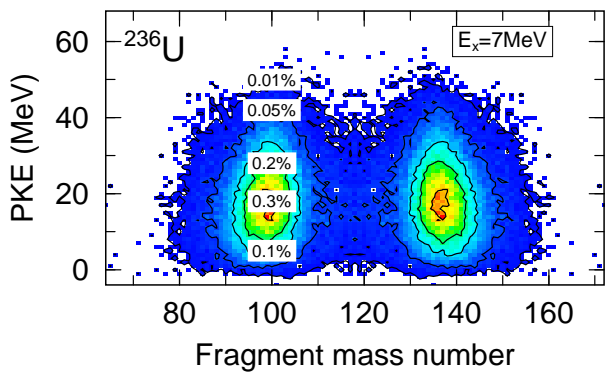}%
\caption{\label{Fig:PKE}Contour map of the prescission kinetic energy as a function of mass number 
of fission fragments at $E_x=7$ MeV for $^{236}$U.}
\end{figure}

In Fig.~\ref{Fig:PKE} we also present the distribution of the prescission kinetic energy (PKE),
the collective kinetic energy in the elongation directionat at the instance of neck rupture.
It is the advantage of dynamical theories like ours to be able to obtain this physical quantity.
As a matter of fact, the TKE shown in Fig.~\ref{Fig-05} is expressed in our model as the sum of the PKE and 
Coulomb repulsion energy between point charges at the scission point.
The difference between the Coulomb repulsion energy with the point-charge model and that with the TCSM which takes into account of the spatially extended diffuse charge distributions of two fission fragments, is negligibly small 
around the scission point and after scission. 
For the details of the Coulomb calculation of the TCSM, see the ref.~\cite{iwam76} and the reference therein. 
Therefore, we used the point charge model for simplicity in this paper.

The PKE 
contains a memory of dynamical nuclear motion from the initial position to scission.
The mean PKE is obtained to be about 18.56 MeV at $E_x=$7 MeV in the present 4D model,
and is almost independent of the fragment mass.
The value amounts to about 10\% of the 
TKE shown in Fig.~\ref{Fig-05}; thus, the PKE is an important component of the TKE of fission fragments.

The reason for the present high average PKE value can be interpreted as follows.
In the Langevin model, there exist various fission paths on the potential energy surface
because each trajectory is affected by the random force, which leads to 
variation of the collective momenta as well as the potential energy gradient event by event.
This leads to, even for the symmetric fission component, 
strong variations of the history of the fission paths and also scission shapes that lead to a broad distribution of PKE, from 0 to around 35 MeV,  
unlike the scission-point model~\cite{Wilkins+1976} which has a fixed value of PKE.  The high component of the PKE values pushes up 
the average PKE value in our model.
However, we recognize that there is still room for discussion of validity of the high average PKE value obtained in this work.
The remarkable agreement of TKE with experimental data as shown in Fig.~\ref{Fig-05}, on the other hand, strongly
indicates that the prescission dynamics described in our 4D Langevin model, hence, 
our PKE distribution, reflects a certain aspect of the correct fission mechanisms.

\begin{figure}[ht]
\includegraphics[width=8cm]{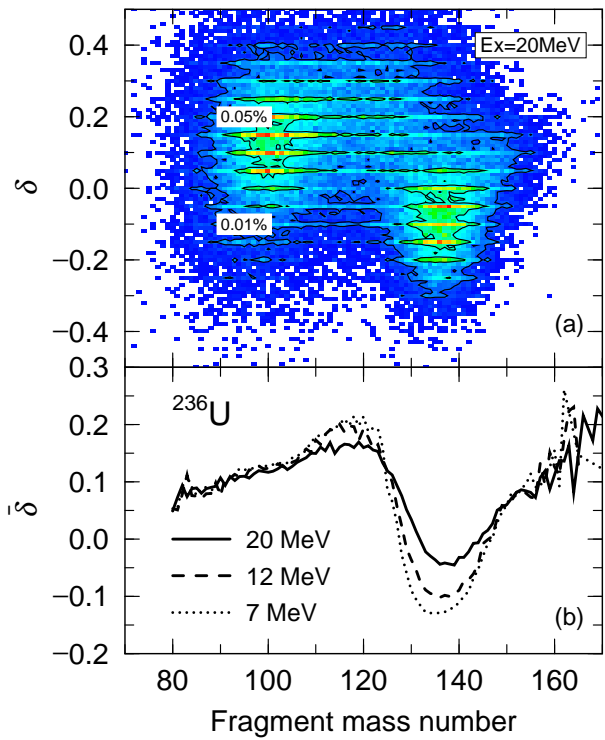}
\caption{(color online) The distribution of the deformation parameter $\delta$ in its dependence on the mass number is illustrated in panel (a)($E_x=20$MeV), while mean values of $\delta$ at excitation energies of 7, 12 and 20 MeV are shown as functions of fragment mass number in panel (b). Both are for $^{236}$U.
}\label{Fig-02}
\end{figure}

In order to investigate the shapes of a fissioning nucleus at the scission point,
we have plotted in Fig.~\ref{Fig-02}~(a) the distribution of the parameter $\delta$ as function of fragment mass number calculated within the 4D TCWS model. 
The light fragments apparently show different deformation from the heavy fragments. 
This feature cannot be achieved in the 3D formulation.

%
\begin{figure}[h]
\includegraphics[width=8cm]{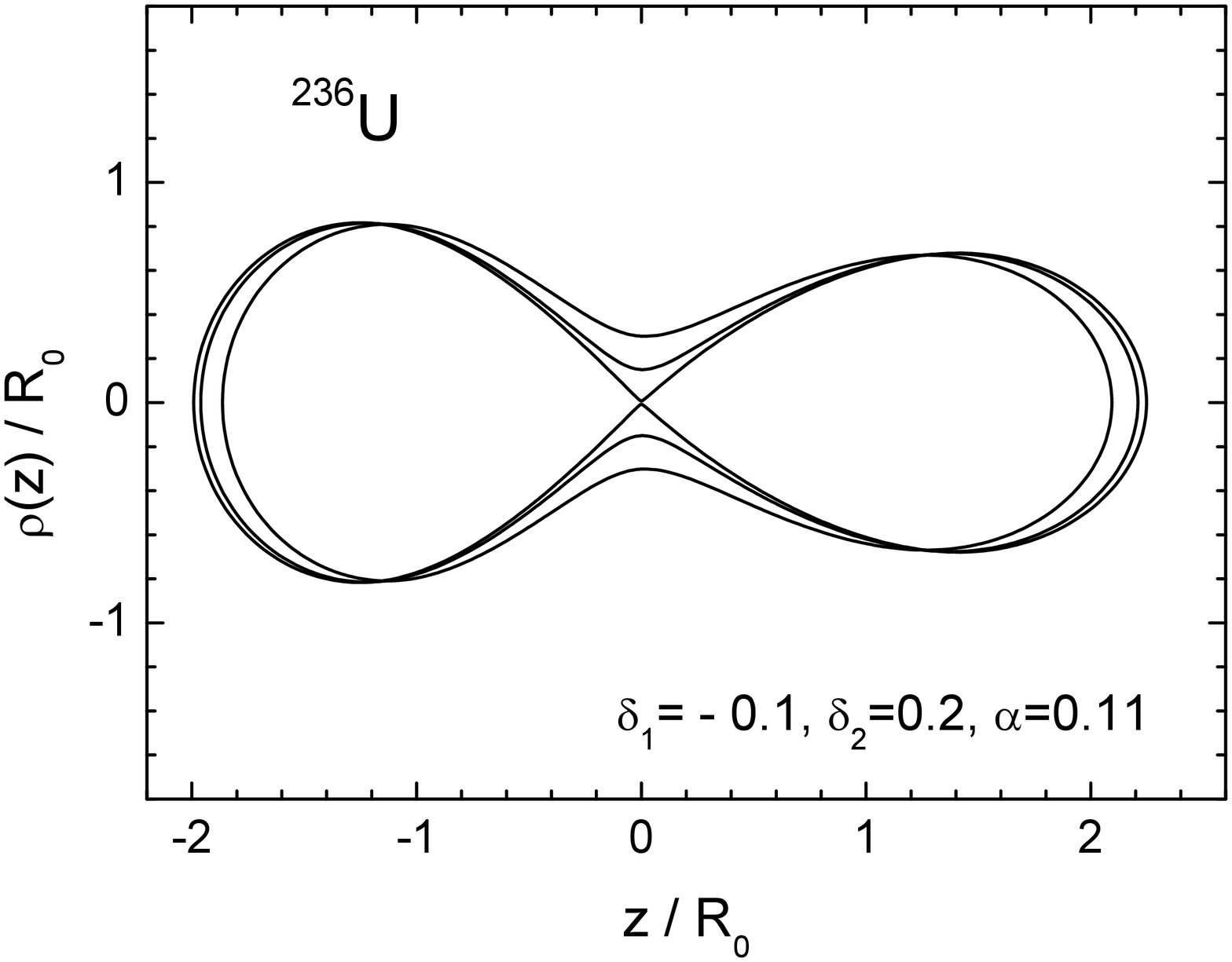}
\caption{\label{Fig:A132-shape}
The average nuclear shape near the scission point. The three curves correspond to $r_{neck}$ = 0, 1 and 2 fm. The $\delta_1$ in all three cases is negative, $\delta_1$= - 0.1. The mass number of heavy fragment is equal to 132. 
}
\end{figure}
Note that the shape of heavy fragment at $A_H=132$ 
has a
negative $\delta$ 
on the average
 but this does not necessarily mean the oblate shape,  
as is seen in Fig.~\ref{Fig:A132-shape}.
The parameter $\delta$ specifies only the deformation of 
the outer part ($z\leq z_1$ or $z_2\leq z$) of the fissioning nucleus.
It can be seen that a nearly spherical 
or even slightly prolate nucleus is produced at $A-H=132$ while
the light fragment is very elongated.

The similarity between the fragment deformation immediately after scission and the neutron multiplicity was already noticed
in the pioneering study by Wilkins~\cite{Wilkins+1976}.
In Fig.~\ref{Fig-02}~(b), the mean value of the deformation $\overline{\delta}$ is shown for three excitation energies.
From the mean values of the deformation $\overline{\delta}(A)$,
we see that on average 
the lighter fragments have more elongated (prolate) shapes  compared to the heavier fragments.
The mean deformation $\overline{\delta}$ also reveals another specific feature, a sawtooth structure
which is remarkably similar to that of the prompt neutron multiplicity $\nu(A)$.
For neutron energies 0.05 MeV $\leq E_n \leq$ 5.55 MeV 
impinging on $^{235}$U, the prompt neutron multiplicities for lighter fragments are independent of the excitation energy $E_x$,
while those for heavier fragments increase as $E_x$ \cite{muller1984} increases. 
The mean deformations $\overline{\delta}$ in our 4D-model show a similar energy dependence as for the prompt
neutrons, i.e.,
the $E_x$ dependence can be seen only in the heavier mass components around $130<A<150$.

\section{ Summary}
We have developed a 4-dimensional Langevin model to improve the description on fission dynamics at low excitation energy
and applied it to fission of $^{236}$U
at low excitation energy, which is a compound nucleus
in neutron induced fission of $^{235}$U.  This system has the most abundant set of experimental information among neutron induced fission to verify the model, and also it is important from the application point of view.
Our model deals with not only the independent deformation
parameters of fission fragments but also with the modifications of the potential such as the infinite-depth
two-center shell model (TCSM) potential and the finite-depth two-center Woods-Saxon (TCWS) potential.
It turns out that the width of the peak of fission fragment mass distributions (FFMDs) in the 4D model
is narrower compared with the 3D model due to the deeper potential valley as a consequence of taking into account the additional degree of freedom.
In spite of this behavior of FFMDs, we have successfully reproduced the experimental total kinetic energy (TKE), which is a good indicator of the nuclear shape 
at scission.
This agreement gives a support to our 4D model, which is ascribed to dealing with independent deformations between the two fragments at scission.
In our model, we predict that about 10\% of the TKE came from the prescission kinetic energy (PKE).

It should be stressed that the present 4D Langevin model reproduces  
the mass distributions of fission fragments, the dependence of the total kinetic energy on the fragment mass, and its standard deviation as a function of the neutron kinetic energy 
(excitation energy of compound nucleus) simultaneously with better accuracy than the 3D Langevin models.
We also find a 
strong correlation between the mass-dependent deformation of fragments at the scission point and the sawtooth structure of prompt neutron multiplicity~\cite{Ivanyuk+2017}
including their dependence on excitation energy.

In the present work, we concentrated our attention on the system of $^{236}$U as the compound nucleus.  It is worth also to apply the present model to, at least, neighboring actinides to see how it can describe fission observables of other nuclei in a systematic manner.  A quantitative analysis of the neutron multiplicity data within the framework of the present 4D model is also promising.  
Application of the linear response theory to calculate the 4-dimensional transport coefficients in a microscopic manner is also a necessary step for refinement of the theory as well as extension to the 5D dynamical model where the $\epsilon$ parameter enters the category of dynamical variables.  All these features are currently under investigation in our group.

\acknowledgments
This work was supported by
the grant ``Comprehensive study of delayed-neutron yields for accurate
evaluation of kinetics of high-burn up reactors"
entrusted to Tokyo Institute of Technology by the Ministry of Education, Culture, Sports, Science and Technology of Japan (MEXT), and 
"Development of prompt-neutron measurement in fission by surrogate reaction method and
evaluation of neutron-energy spectra" entrusted to Japan Atomic Energy Agency by MEXT.  
The authors also thank the WRHI (World Research Hub Initiative) program of Tokyo Institute of Technology, and 
IAEA CRP on beta-delayed neutrons (F41030).


\end{document}